# Tunable Semimetallic State in Compressive-strained SrIrO$_3$ Films Revealed by Transport Behaviors

Lunyong Zhang [1]*, Qifeng Liang[2], Ye Xiong[3], Binbin Zhang[1], Lei Gao[4], Handong Li[4], Y.B Chen[5]*, Jian Zhou[1], Shan-Tao Zhang[1], Zheng-Bin Gu[1], Shu-hua Yao[1]*, Zhiming Wang[4], Yuan Lin[4], and Yan-Feng Chen[1]

[1] *National Laboratory of Solid State Microstructures & Department of Materials Science and Engineering, Nanjing University, 210093 Nanjing, P. R. China*

[2] *Department of Physics, Shaoxing University, 312000, Shaoxing, P. R. China*

[3] *College of Physical Science and Technology, Nanjing Normal University, 210097, 210093, P. R. China*

[4] *State Key Laboratory of Electronic Thin Films and Integrated Devices, University of Electronic Science and Technology of China, Chengdu 610054, China*

[5] *National Laboratory of Solid State Microstructures & Department of Physics, Nanjing University, 210093 Nanjing, P. R. China*

Orthorhombic SrIrO$_3$ is a typical spin-orbit-coupling correlated metal that shows diversified physical properties under the external stimuli. Here nonlinear Hall effect and weakly temperature-dependent resistance are observed in a SrIrO$_3$ film epitaxially grown on SrTiO$_3$ substrate. It infers that orthorhombic SrIrO$_3$ is a semimetal oxide. However, linear Hall effect and insensitive-temperature-dependent resistance are observed in SrIrO$_3$ films grown on (La,Sr)(Al,Ta)O$_3$ (LSAT) substrates, suggesting a tunable semimetallic state due to band structure change in SrIrO$_3$ films under different compressive strain. The mechanism of this evolution is explored in detail through strain-state analysis by reciprocal space mapping and electron diffraction, carrier density and mobility calculations, as well as electronic band structure evolution under compressive strain (predicted by tight-binding approximation). It might suggest that the strain-induced band shift leads to the semimetallic tuning in the SrIrO$_3$ film grown on from SrTiO$_3$ to LSAT substrates. Our findings illustrate the tunability of SrIrO$_3$ properties and pave the way to induce novel physical states in SrIrO$_3$ such as the proposed topological insulator state in heterostructures.

## I. INTRODUCTION

Recent years, the research of transition-metal oxides has been extended to the 5d orbit materials that show a series of novel physics phenomena, such as spin-orbit coupling assistant Mott insulator, Weyl semimetal and correlated topological insulator. These phenomena are attributed to the increased spin-orbit coupling (SOC) in 5d transition metal that can compete with electron correlation (U) and/or crystal field energy ($\Delta$). A representative system of these materials is the iridate $Sr_2IrO_4$ whose SOC is about 0.4eV, comparable to U (~0.5eV) and $\Delta$ (~2eV)[1,2]. It induces a half filled $J_{eff}$ =1/2 Mott state in $Sr_2IrO_4$ that is a prototypical material of SOC assistant Mott insulator. The study of $Sr_2IrO_4$ inspires a lot of theoretical study about the physical properties under the synergetic effects of electron correlation and SOC.

Experimentally, some of work are focused on orthorhombic $SrIrO_3$ [SIO, the three-dimensional member of the Ruddlesden-Popper series of iridates $Sr_{n+1}Ir_nO_{3n+1}$ (n=1, 2, $\infty$)]. In $SrIrO_3$, a large electron hopping leads to weak electron correlation U, which makes the $SrIrO_3$ a good playground to exploring phase diagram with SOC dominant. Some reports claim that SIO is a paramagnetic correlated metal, different from most of iridates as insulating conduction and antiferromagnetism [1,3]. As to the band structure of SIO, varied theoretical pictures have been proposed. The early calculation based on LDA+U+SOC showed SIO is just a simple metal with a small hole-packet at the highest symmetry point of the Brillouin zone[1]. However, recent theoretical studies demonstrate an electron pocket coming from a line node and a hole pocket slightly crossed by Fermi level in the band structure of SIO [4,5], implying that SIO might be a semimetal (seemly verified by the recent ARPES studies[6]) and may be modified into a strong topological insulator in some heterostructures. In addition, the hole-like band in the semimetal state of SIO is nearly flat in a wide *k* space at the Fermi level[5]. It leads to a prediction that the hole pocket might be easily removed by external stimuli (such as strain effect, electron doping). Tunable semimetallic state in SIO in turn may be detected. Experimentally, Liu, *et.al,* has reported that compressive-strain enhances the asymmetry of the electron-hole carrier mobility in SIO films. And Biswas, *et al*, recently claims that compressive strain could increase the electron-correlation in SIO film [7], indicating a band width decrease under compressive stain. However, tensile strain weakly affects the electron-hole mobility symmetry [8]. Nevertheless, no distinct experimental results show the semimetal ground state and tunable semimetallic state in SIO.

Inspired by forgoing works, we have an intuitive understanding for SIO. The electronic band structure of $Sr_{n+1}Ir_nO_{3n+1}$ series consist of the particular $J_{eff}$=1/2 and $J_{eff}$=3/2 bands originated by the delicate interplay between SOC and correlation [2]. These results suggest that the band structure of SIO is sensitive to compressive strain. Here we detected Hall effect evolution from nonlinear to linear in SIO films with lattice compressive strain increased from -1.1% to -2.1%, confirming the semimetal ground state and indicating a tunable semimetallic state in SIO. The underlying physical mechanism is explored through electron transport, magneto-transport, valence band spectra and the electronic structure calculations by the tight-binding approximation Hamiltonian.

## II. EXPERIMENTAL DETAILS

The largest compressive strain case considered by Liu *et al.* [8] was provided by (001)-$SrTiO_3$ substrate. It is relatively small and can not adjust the semimetal state of SIO. In the present paper, we extended the compressive strain through the (001)-$(LaAlO_3)_{0.3}(Sr_2AlTaO_6)_{0.7}$ (LSAT) substrate to further exploring the effect of large compressive strain on the semimetal state of $SrIrO_3$ film. Our recent transmission electronic microscopy (TEM) study has shown that the main orientation relationship in $(SIO)_o$/(001)-STO is $(SIO)_o$[001]//STO[010] and $(SIO)_o$ [$\bar{1}$10] //STO[100] [9], and we believe this orientation relationship to be still validate in $(SIO)_o$/(001)-LSAT since the substrates have same rectangular lattice structures with small differences in lattice parameter in the (001) plane. Theoretically, the average stains in fully strained $SrIrO_3$ films are therefore -1.1% on the STO substrate and -2.1% on the LSAT substrate, corresponding to the (110)-plane lattice parameters $a_x$=7.89Å and $a_y$=7.92 Å for SIO, the (001)-plane lattice parameters $a_x$=$a_y$=3.9034

Å for STO and $a_x=a_y=3.868$ Å for LSAT.

The films were synthesized in 25Pa oxygen atmosphere through pulsed laser deposition (PLD) at 750 °C substrate temperature and 10/3 Hz laser pulse frequency. Detailed synthesis process can be consulted in Ref. 9. Before films synthesis, the STO substrates were chemically treated and annealed to obtain stepped surfaces according to the method reported in Ref. 10. The LSAT substrates were etched in a 0.2mol/L NaOH solution for 1 min and then ultrasonic washed in deionized water for 5 min to clean the surfaces. The LSAT substrates were not annealed here since we found it is difficult to obtain a stepped LSAT surface with low roughness suitable for film growth through annealing.

The $\theta$-$2\theta$ scanning and reciprocal space mapping (RSM) were carried out on a Bede D1 high resolution X-ray diffractometer (XRD) with Cu-K$\alpha$ ray. Film surface morphology was imaged through an Asylum Cypher atomic force microscopy (AFM) system. The microstructure and strain state of SIO films were characterized by a Tecnai F20 transmission electronic microscope (TEM). The TEM specimens were prepared by conventional method including grinding, polishing and ion-milling. Electron transport measurements were carried out on a Quantum Design physical property measurement system (9T-PPMS) through the four-probe method. The valence band spectra were measured on a Thermo Scientific K$\alpha$ X-ray Photoelectron Spectrometer (XPS) system at room temperature. Before collecting the valence band spectra, the surfaces of the samples were *in-situ* cleaned for 40s using 1000eV ion beam which was proved to not change the film stoichiometry by the core level peak integration of the elements obtained by XPS analyses[11]. The electronic band structures with and with compressive strain were obtained by a tight-binding Hamiltonian proposed by Carter to explain the adjustable semimetal behavior of SIO[4].

**III. RESULTS AND DISCUSSIONS**

Fig. 1 summarizes the structure and strain-state characterization of the SIO films grown on STO and LSAT substrates. Fig. 1(a) and (b) depict the $\theta$-$2\theta$ scanning patterns of SIO films grown on STO and LSAT substrate, respectively. One can see that there are (110) and (220) peaks belonging to the SIO films. And there is no any reflection coming from impurity phase. Fig. 1(c) and (g) show the morphology of the SIO films, revealed by TEM, grown on STO and LSAT substrates, respectively. These images substantiate that the SIO films has quite uniform thicknesses (thickness ~7 nm), and there is no impurity phase. Fig. 1(d) is the selected area electron diffraction pattern taken from area covering both SIO film and STO substrate. The electron beam was aligned along [1,-3, 0] zone axis of STO. As shown in this pattern, reflections coming from film and substrate have clear splitting along growth direction (see right inset, it is (004) pole of STO), however, there is no splitting parallel substrate surface (see upper left inset, it is (3,1,0) pole of STO). We thus can conclude that SIO film is coherently strained on STO substrate. AFM images display the terrace-like structure on the surfaces of SIO film grown on both STO substrate (see Fig.1 (e)], and height between neighboring terraces is roughly 4 Å. It infers that the film was grown by layer-by-layer mechanism, generally suggesting a high quality epitaxial film. RSMs around the pseudo-cubic (013) reflection of the STO substrates [see Fig. 1 (f)] reveal distinct elongated SIO spots along the out-plane direction. Elongation of SIO reflection infers that the film is quite thin (crystal truncation rod). Of particular importance is that as the RSMs indicates, the SIO film is in plane coherently constrained by the substrates along substrate surface directions. This is in agreement with electron diffraction result (see Fig. 1(d)). The strain state and surface morphology of SIO grown on LSAT substrate (see Fig. 1(g)-(j)) are quite similar to those of SIO grown on STO substrate. One minor difference is that Fig.1 (j) displays a slightly split spot assigned to the LSAT substrate, which implies there might be some lattice distortions deviating from the perfect (001)-LSAT substrate in crystalline orientation. Even so, the SIO reflection spot is narrow and is matched with the main spot of the substrate [Fig.1 (j)]. It is worthwhile to emphasize that the SIO films are fully strained on both STO and LSAT substrates along in-plane directions, their compressive strains are -1.1% on the STO substrate and -2.1% on the LSAT substrate, respectively.

To clearly compare the transport between the films grown on STO and LSAT substrates, we plotted the temperature-dependent resistance curves in Fig. 2(a). The corresponding resistivity at the low temperature limiting ~2K is about 2 mΩ cm and 6 mΩ cm for the SIO/STO film and the SIO/LSAT film, respectively, in agreement with the resistivity in order reported in the 20 nm thick SIO films on STO and LSAT substrates [12]. As shown in Fig. 2(a), compared with the SIO film grown on STO, the SIO film on LSAT demonstrated a trend to approach semiconductor or insulator. The SIO film on LSAT also showed a shallow resistance minimum at about 150 K, indicating a metal-insulator-like transition (MIT). The Kondo effect can be ruled out in this sample because the logarithmic temperature-dependent resistance relation is not held. Furthermore, there are no magnetic impurities in the SIO films. Detail analyses confirm that the semiconductor state resistivity of the films on both the substrates can not be well fitted by common resistivity temperature relation models for semiconductor including the thermal activation model[13] and the three dimensional weak localization model [14].

For ultrathin films exhibiting MIT, the two dimensional weak localization is a widely proved mechanism to trigger the transition, whose conductance linearly depends on the logarithm of temperature and can be modeled as [14]

$$G_{2D}(T) = G_0 + p\frac{e^2}{\pi h}\ln\left(\frac{T}{T_0}\right) \quad (1)$$

where $\tau_i$ is inelastic scattering relaxation time and $\tau_e$ is elastic scattering relaxation time, $G_0$ is the Drude conductance, $a$ and $T_0$ are constants, $e$ is electron charge and $h$ is the Planck constant, and $p$ is the temperature exponent of inelastic scattering length $l_i \sim T^{p/2}$. $p=3$ for electron-boson scattering and $p=1$ for electron-electron scattering [14]. As demonstrated in Fig. 2c, both of the conductance of the two sample films bear good linear dependence on ln$T$ in certain low temperature region. For the SIO film on LSAT, the linear dependence could be well fitted by Eq. (1) with $p$ of 0.98, implying that the electron-electron scattering dominates electric conductivity. The fitting $p$ is about 1.42 for the SIO film on STO, which means that the carriers are scattered by both electron-electron and electron-boson interaction. These results are in agreement with our previous work[15] that substantiates resistance minima induced by weak localization in thick SIO films grown on STO substrates. Therefore, weak localization is still a dominated mechanism of the conduction in both SIO/LSAT and SIO/STO film.

It should be emphasized that electron-electron scattering is dominant in the SIO/LSAT film, consistence with the report of Biswas *et al.* that both electron-correlation and disorder can not be ignorable in the MIT of SIO films under large compressive strain[7]. Enhancement of electron-electron correlation can be understood as the decreased band-width $W$ with increasing compressive strain in the (110) plane of SIO film. Quantitatively, $W$ is proportional to $l^{-3.5}\sin(\theta/2)$ for a ABO$_3$ perovskite compound [16]. Given that the Ir-O-Ir angle $\theta$ of SrIrO$_3$ would decrease under increased compressive strain[17] and the Ir-O length $l$ would not be affected dramatically for example the observations in Sr$_2$IrO$_4$ film[18, 19]. $W$ thus should be decreased by increasing compressive strain, which in turn causes the enhanced effective correlation $U/W$.

Close inspection of Fig. 2(a) indicates that the resistivity of both two SIO films is varied smaller than 1.4 when the temperature ranges from 300K to 2K. It is a hint suggesting the SIO might be a semimetal [8]. In principle, a semimetal system would simultaneously have electron and hole carriers which can compensate to keep a relatively temperature-insensitive electrical resistance. The conduction of a semimetal could always be described by the two-carrier transport model [20] which states that the Hall effect is nonlinear for a semimetal in most cases. As shown in Fig. 3(a) and (b), the Hall effect measurements of our samples show obviously nonlinear Hall traces for the SIO films grown on STO substrate (Fig. 3(a)), but the Hall traces are roughly linear for the SIO films grown on LSAT substrate (Fig. 3(b)). The Hall coefficients of them are negative, indicating that electron is the major carriers. Quantitatively, we adopted the two-carrier transport model for an isotropic material [20], to simulate the temperature-dependent-resistance and magnetic-field dependent Hall coefficient:

$$\rho = \frac{1}{e(|n_1|\mu_1 + |n_2|\mu_2)} \quad (2)$$

$$R_{\mathrm{H}}(B) = \frac{\left(n_1\mu_1^2 + n_2\mu_2^2\right) + \left(\mu_1\mu_2 B\right)^2\left(n_1 + n_2\right)}{e\left[\left(|n_1|\mu_1 + |n_2|\mu_2\right)^2 + \left(\mu_1\mu_2 B\right)^2\left(n_1 + n_2\right)^2\right]} \tag{3}$$

where $n_i$ (i=1, 2) is carrier concentration, and it is negative for electron and positive for hole. $\mu_i$ is carrier mobility, $B$ is magnetic field. It can be seen $R_{\mathrm{H}}$ is intensively magnetic field dependent when electron and hole carriers, comparable in concentration and mobility, simultaneously exist in a system according to Eq. (3), leading to a strongly nonlinear Hall resistance trace. And $R_{\mathrm{H}}$ is magnetic-field independent for a system with equal electron and hole in concentration, SIO therefore should not be a simple $J_{\mathrm{eff}}=1/2$ system with equal concentrations of electron and hole (as assumed in Ref.[8]) owning to the observed nonlinear Hall effect of the SIO/STO films; also $R_{\mathrm{H}}$ would be slightly magnetic field dependent when the difference of the mobility or/and the density is extremely large for the electron and the hole carriers, this would cause a nearly linear Hall trace. Practically, there are two simplified methods to extract the carrier mobility and the carrier density in a semimetal on the basis of Eq. (2) and (3) for overcoming the difficulty in full magnetic field fitting of the $R_{\mathrm{H}}(B)$ under the constraint of Eq.(2). The first method assumes that the electron carrier and the hole carrier have same concentrations (equal density method); the second one does same mobility for both hole and electron (equal mobility method) [21]. Since Hall trace is linear for a system with equal carrier densities of electron and hole, the equal density method is obviously not applicable for our SIO/STO sample film in line to Eq. (3). For the SIO film on LSAT, its carrier density and mobility have been calculated through the one electron band model, the equal density method as well as the equal mobility method since its Hall is linearly dependent on magnetic field [Fig.3(b)]. It was shown that the one electron band model gave out electron density far larger (about 600-700 times) than that in the SIO/STO film obtained through the equal mobility method. It is not explicitly reasonable. In SIO/LSAT system, both equal density and equal mobility method yield quite similar carrier concentrations and mobility. Therefore, we simulate data of SIO/LSAT and SIO/STO systems by means of equal mobility method.

Under the equal mobility assumption, Eq. (2) can be rewritten as $\rho(B=0)=[e\mu(n_{\mathrm{h}}-n_{\mathrm{e}})]^{-1}$

with $\mu=\mu_e=\mu_h$, and Eq.(3) in low field limit can be simplified as $R_H = (n_e+n_h)/[e(n_h-n_e)^2]$. In line to these relations, $n$, $\mu_e$ and $\mu_h$ can be extracted out, and they were plotted at different temperatures in Fig. 3(e) and Fig. 3(f). The carrier mobility is decreased with temperature increasing, a common feature of mobility. It is about 100~200 cm$^2$ V$^{-1}$s$^{-1}$ in the SIO/STO film, 5~8 times lower than that of the film grown on LSAT substrate. The carrier density is slightly increased with temperature, similar with that of bismuth semimetal film[21]. Quantitatively, the densities both of the electron and hole in the SIO/STO film are $10^{19}$ cm$^{-3}$ in order of magnitude, in agreement with the magnitude in a 18nm thick SIO/STO film and comparable with that of typical semimetals such as graphene [8]. *The carrier density in the SIO/LSAT film is about $(4~12) \times 10^{17}$ cm$^{-3}$, about 30~50 times smaller than that in the SIO/STO film.*

Defining the asymmetry of carrier density as $\kappa=(|n_e|-n_h)/(|n_e|+n_h)$, Fig.3 (f) demonstrates that $\kappa$ is $10^{-4}$ to $10^{-3}$ in order of magnitude in the films, a typical feature of semimetal. The SIO/LSAT film has nearly zero $\kappa$, independent on temperature, means roughly equal density of electron and hole, corresponding to its nearly linear Hall traces. In the SIO/STO film, $\kappa$ is about 0.005 at 5K and increases to about 0.01 at 8K, evidently larger than that in SIO grown on LSAT substrate. The carrier asymmetry leads to a detectable nonlinear Hall traces in SIO grown on STO substrate.

We also studied the magneto-conductivity behaviour of SIO films grown on both STO and LSAT substrates. Fig. 3(c) and (d) display the magnetic-field-dependent MC of SIO films grown on STO and LSAT substrates, respectively. As shown in Fig. 3(c) & (d), the general feature of magnetic-field dependent MC is the cusp-like curve in these figures. The same magneo-conductivity behaviour is also observed in the typical semimetallic bismuth thin film. The cusp-feature is attributed to competition between SOC and weak localization. In details, the time-reversal electron path in weak localization is destroyed by a uniform magnetic field, which in turn reduces positive MC. It competes with strong SOC that can cause negative MC at low magnetic-field range. Quantitatively, this phenomenon can be described by Hikami-Larkin-Nagaoka (HLN) equation:

$$\frac{\Delta G}{G_{u}} \propto \left\{ \psi\left(\frac{1}{2}+\frac{B_{e}}{B}\right) - \psi\left(\frac{1}{2}+\frac{B_{i}+B_{SOI}}{B}\right) + \frac{1}{2}\left[\psi\left(\frac{1}{2}+\frac{B_{i}}{B}\right) - \psi\left(\frac{1}{2}+\frac{B_{i}+2B_{SOI}}{B}\right)\right] \right\} \quad (4)$$

where $\psi(x)$ is the digamma function, $\Delta G = G(B)-G(B=0)$ and $G_u = e^2/(\pi h) \approx 1.2 \times 10^{-5}$S is a universal value of conductance. $B_e$, $B_i$ and $B_{SOC}$ are the equivalent fields of elastic scattering, inelastic scattering and the scattering induced by SOC, respectively. We can extract the SOC strength by fitting the traces with the HLN model [22]. The fitting results demonstrate that the magnitude of $B_e$ is certainly much larger than that of $B_i$ and $B_{SOC}$, at least 100 times, the condition $\tau_i \gg \tau_e$ mentioned in Eq. (1) does hold in our SIO films. Of more important feature is that $B_{SOC}$ is decreased about 80% in average with the compressive strain increasing from -1.1% to -2.1% (see Fig. 3(h)), this is in agreement with the large mobility different between SIO grown on STO in contrast to SIO grown on LSAT (in definition $B_{soc} = \frac{\hbar}{4eD\tau_{SOC}}$ and $D$ is diffusion constant and is proportional to carrier mobility) [see Fig. 3(e)]. And it should be mentioned that a low magnetic-field region, the magnetoresistance is proportional to $B^2$ that is universally observed in conventional semimetals (like bismuth thin films).

We further study the semimetallic state evolution in the SIO films through valence band XPS spectra as shown in Fig. 4. Obviously, both of the valence XPS spectra show finite counts at Fermi level. But the count of the SIO/STO film is much larger than that of the SIO/LSAT film, suggesting a higher density of state in the film on STO, in agreement with the result of carrier density calculation (see Fig. 3(f)).

To fully understand the tunability of semimetal behaviour in SIO films under the compressive strain, we calculate the electronic band structure without and with compressive strain [shown in Fig. 5 (a)-(d)] of SIO by means of a tight-binding approximated Hamiltonian proposed by Carter et al.[4] The Hamiltonian is written as:

$$\begin{aligned}
H = &\lambda_k + \mathrm{Re}\left(E_k^q\right)\gamma_x + \mathrm{Im}\left(E_k^q\right)\sigma_z\gamma_y + \left(E_k^z\right)\gamma_x \\
&+ \mathrm{Re}\left(E_k^d\right)v_x\gamma_x + \mathrm{Im}\left(E_k^d\right)v_y\gamma_y \\
&+ \mathrm{Re}\left(E_k^{qo}\right)\sigma_y v_z\gamma_y + \mathrm{Im}\left(E_k^{qo}\right)\sigma_x v_z\gamma_y \\
&+ \mathrm{Re}\left(E^{zo}\right)\sigma_y v_y\gamma_z + \mathrm{Im}\left(E_k^{zo}\right)\sigma_x v_y\gamma_z \\
&+ \mathrm{Re}\left(E_k^{do}\right)\sigma_y v_y\gamma_x + \mathrm{Im}\left(E_k^{do}\right)\sigma_x v_y\gamma_x
\end{aligned} \quad (5)$$

In this model Hamiltonian, only direct $J_{\mathrm{eff}}=1/2$ orbits hoppings along the nearest and next-nearest neighbor Ir-Ir pair are taken into account. The superscripts $q$, $z$, and $d$ indicate hopping between Ir atoms within a (001) layer, between layers, and between next-nearest neighbors, respectively, and $o$ represents hopping between $J_{\mathrm{eff}}^z = 1/2$ and $-1/2$. $\sigma$ is the Pauli matrices. $\gamma$ Pauli matrices for the (001) in-plane sublattice, $v$ is Pauli matrices for the (110) or ($1\bar{1}0$) in-plane sublattice. The coefficients in Eq. (5) are

$$\begin{aligned}
\lambda_k &= \lambda + t_{xy}\cos(kx)\cos(ky), \\
E_k^q &= (2t - it')[\cos(kx) + \cos(ky)], \\
E_k^z &= 2t\cos(kz), \quad E_k^{zo} = (1-i)t_z^o\cos(kz), \\
E_k^d &= t_d[\cos(kx) + \cos(ky)]\cos(kz) \\
&\quad + it_d'[\sin(kx) + \sin(ky)]\sin(kz), \\
E_k^{qo} &= [t_{1q}^o\cos(kx) + t_{2q}^o\cos(ky)] \\
&\quad - i\left[t_{2q}^o\cos(kx) + t_{1q}^o\cos(ky)\right], \\
E_k^{do} &= t_d^o[\sin(ky) + i\sin(kx)]\sin(kz),
\end{aligned} \quad (6)$$

where $\lambda$ is the SOC, $t$ nearest neighbor intra-orbital hopping, $t_{xy}$ next-nearest neighbor in-plane hopping of $xy$ orbits, and $t'$ the hopping between 1D orbits in the plane ($xz$ and $yz$). $t_d$ and $t_d'$ are originated from intra-orbital diagonal hopping of next-nearest neighbor atoms from different layers. $t_z^o$ represents the hopping from 1D orbits to $xy$ orbit between two adjacent atoms on different planes, $t_{1q}^o$ and $t_{2q}^o$ come from the nearest neighbor in-plane hoppings between 1D and $xy$ orbital. Finally, $t_d^o$ arises from out-of-plane next-nearest neighbor 1D to $xy$ orbit hopping. More details about this model Hamiltonian could be found in the Ref.[4]. In this work, we intentionally decrease three major hopping parameters $t$, $t'$ and $t_{xy}$ to simulate the effect of compressive strain. It is well known that the Ir-O-Ir bond angle is decreased under in-plane compressive strain; it in turn leads to smaller hoping parameter. One can see from Fig.5(a) that SIO does have semimetallic band structure that is reported by Carter[4] (all parameters are the same as those in reference 4. To make successive discussion clearly, in the origin SIO case, the three major parameters $t$, $t'$ and $t_{xy}$ are taken as -0.6, -0.15 and -0.3, respectively), a line node appears near the U point and provides the electron carriers, two tops of the $(J_{eff}, J_{eff}^z)$=(1/2, -1/2) valence band near the Γ and the S points supply the hole carriers. As depicted in Fig. 5 (a) & (b), the band structure under decrease of either $t$ or $t'$, the semimetallic behavior around Fermi level has not much changed with respect to state without compressive strain. And we do see the band-width narrow with decreasing of $t$ from Fig. 5(a). But decreasing $t_{xy}$ has dramatically modification of semimetallic behaviour at Fermi level (see Fig. 5(c)). We can see that with $t_{xy}$ decreases, hole's branches along U-R are down-shifted from Fermi level. The compressive strain breaks the proportionality between electron and hole that is observed in SIO grown on STO substrate. And electron part is a little over-domination over hole one, which is demonstrate in SIO films grown on LSAT substrate. The energy band structure with all $t$, $t'$ and $t_{xy}$ decreased is shown in Fig. 5(d). The modification of electronic bands is the combination of those shown in Fig. 5(a), (b) and (c). We therefore conclude that compressive strain is one of crucial ingredient to realize the tuning of semimetallic state of SIO films.

**IV.** CONCLUSIONS

In summary, we have studied the semimetallic state tunability of orthorhombic $SrIrO_3$ films under different compressive strain through transport properties analyses. Nonlinear Hall effect, temperature-insensitive resistance and valence band of XPS were observed in the SIO/STO film with small in-plane compressive strain ($\varepsilon$ = -1.1%), unambiguously suggesting that SIO is a semimetal system with both electron carrier and hole carrier. As compressive stain increased to -2.1% (SIO/LSAT film), nonlinear Hall effect disappears and the electron hole density asymmetry decreases from near zero to ~0.84% in the SIO/STO film. The most possible mechanism responsible for the evolution is that compressive strain downshifts the hole's branches at Fermi level.

The results of present work suggests the half filled $J_{\text{eff}}$=1/2 band of SrIrO$_3$ is a semimetal band with equal electron and hole numbers, and its semimetal property can be easily modulated by strain. This in turn leads to multi-functionalities in SIO film and similar correlated Weyl semimetal systems.[21] Studying the semimetallic state evolution of SIO films with continuously varied strain can provide more insight on the issue. For this task, the piezoelectric substrate such as Pb(Mg$_{1/3}$Nb$_{2/3}$)O$_3$-PbTiO$_3$ (PMN-PT) would be a proper candidate [23].

## ACKNOWLEDGEMENT

The authors would like to acknowledge Professor D Vanderbilt and Dr Hongbin Zhang of Rutgers University, Professor Hae-Young Kee of University of Toronto for their insight discussions and suggestions in the modification of the manuscript. We also thank Professor Di Wu of Nanjing University for the help in film growth equipment. This work was financial supported by the Nation Science Foundation of China (11374149, 51032003, 50632030, 10974083, 51002074, 10904092), the New Century excellent talents in University (NCET-09-0451). The author Lunyong Zhang also acknowledges the financial supports of the Nation Science Foundation of China (51402149), the China Postdoctoral Science Foundation (2013M530250) and the Jiangsu Province Postdoctoral Science Foundation (1202001C), as well as the Jiangsu Province Postdoctoral Sponsor Program.

Corresponding author:

Shu-Hua Yao: shyao@nju.edu.cn ;

Y. B. Chen: ybchen@nju.edu.cn ;

Lunyong Zhang: allen.zhang.ly@gmail.com;

**Figures captions**

**FIG. 1** (a) and (b) are XRD $\theta$-$2\theta$ patterns of the SIO films grown on STO and LSAT substrate, respectively. (c) and (d) are low-magnification TEM image and electron diffraction pattern ([1,-3,0] zone axis of STO) taken from area covering both SIO film and STO substrate. (e) is the AFM images of the sample SIO films grown on STO substrate. (f) Reciprocal space mapping (RSM) around the pseudo-cubic (310) reflection of STO substrate. (g) and (h) are low-magnification TEM image and electron diffraction pattern ([1,-3,0]zone axis of LSAT) taken from area covering both SIO film and LSAT substrate, respectively. (i) is the AFM images of the sample SIO films grown on LSAT substrate. (j) Reciprocal space mapping (RSM) around the pseudo-cubic (013) reflection of LSAT substrate. The split spots assigned to the LSAT substrate in (j) are ascribed to some grains with crystalline orientation slightly deviating from the orientation of a perfect (001)-LSAT substrate.

**FIG. 2** Temperature dependent resistance of SIO films grown on STO and LSAT substrates (a) is the raw data of sheet resistance. At high temperature, resistance of SIO/STO film is proportional to temperature. Data of Ref. 11 have also been added for comparison. (b) is the plots of conductance $G$ *vs* ln$T$ according to weak localization.

**FIG. 3** (a) and (b) give out the Hall effect of the SIO/STO film and the SIO/LSAT film, respectively, showing obvious nonlinear variation in (a) and near linear variation in (b). The inset in (a) shows the Hall coefficient variation with $B$. (c) and (d) are the dependence of MR of the films on magnetic field. At low field limit, MR$\propto B^2$ relation held for two-carrier transport system. (e) and (f) plot the temperature-dependent mobility and carrier concentration. In (f), the temperature-dependent density asymmetry in these two samples is also shown. (g) and (h) depict the temperature-dependent effective magnetic field $B_i$ and $B_{SOI}$ of SIO grown on STO and LSAT substrates, respectively. The dots in (a), (b), (c) and (d) correspond to the raw data and the curves are the fitted results.

**FIG. 4** Valence state XPS spectra of the SIO films grown on STO and LSAT substrates.

**FIG. 5** (a)-(c) are the electronic band structures, calculated by tight-bonding approximated Hamiltonian, that show semimetallic electron structure evolution without and with compressive strain simulated as decreasing $t$, $t'$ and $t_{xy}$, respectively. (d) is the electron band structure without and with compressive strain (decreasing all of $t$, $t'$ and $t_{xy}$).

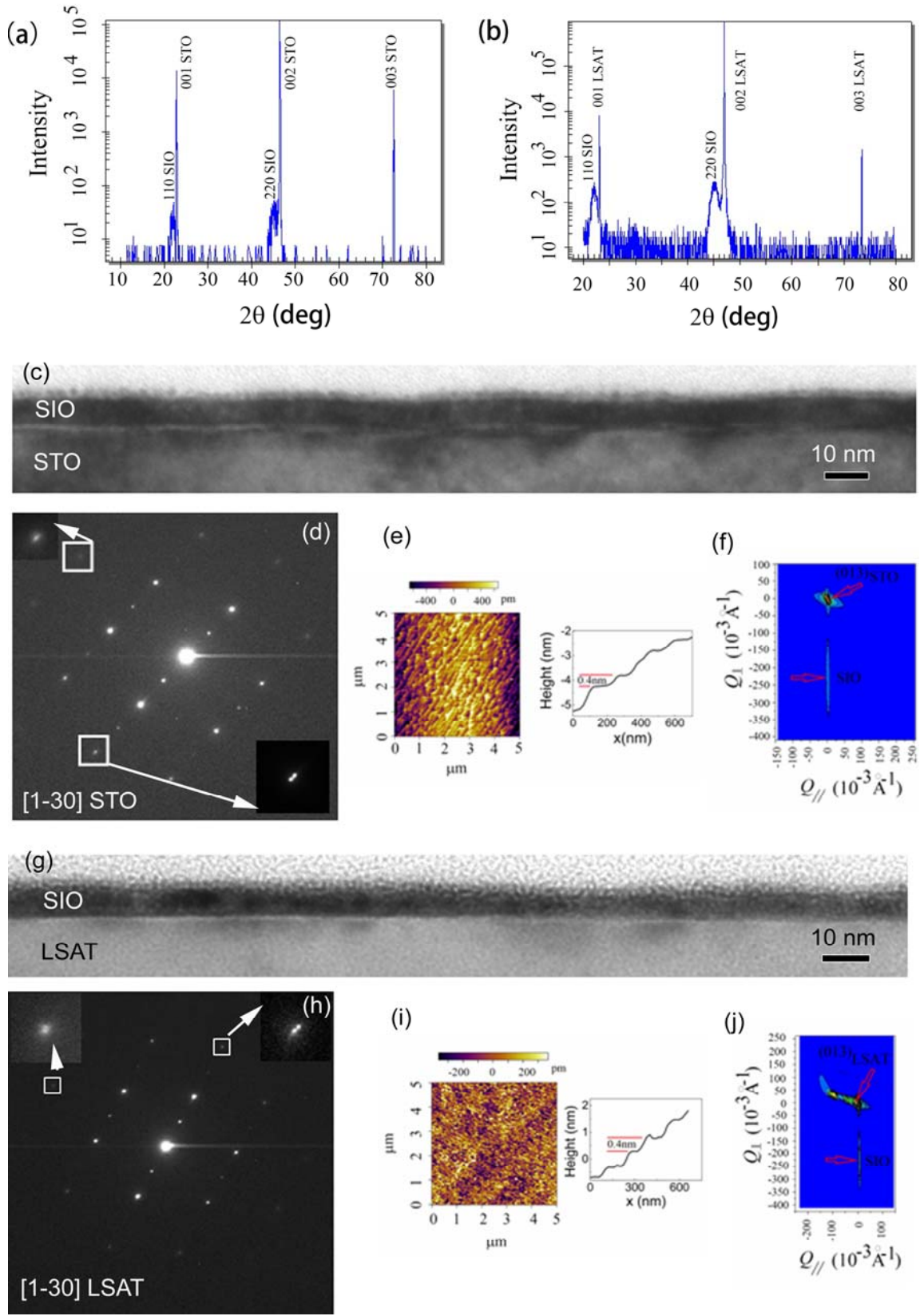

FIG.1

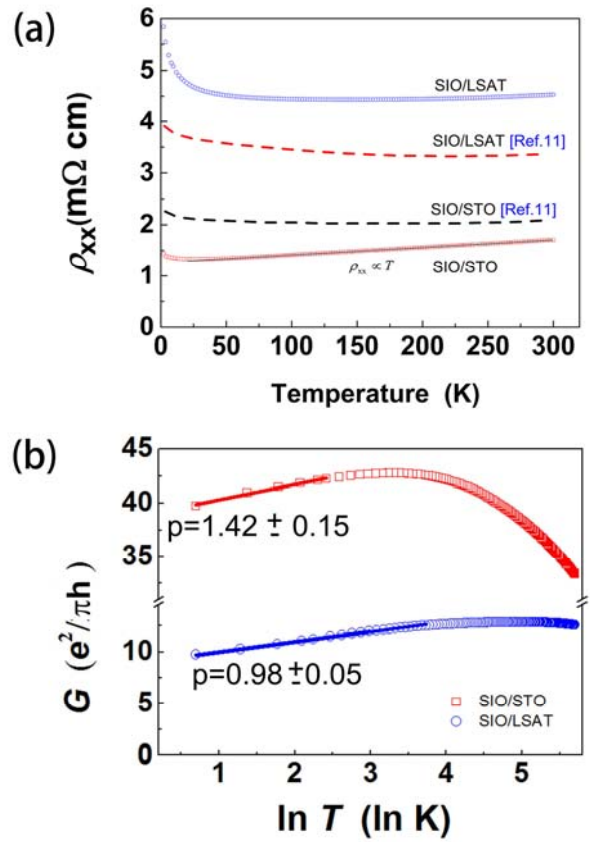

FIG. 2

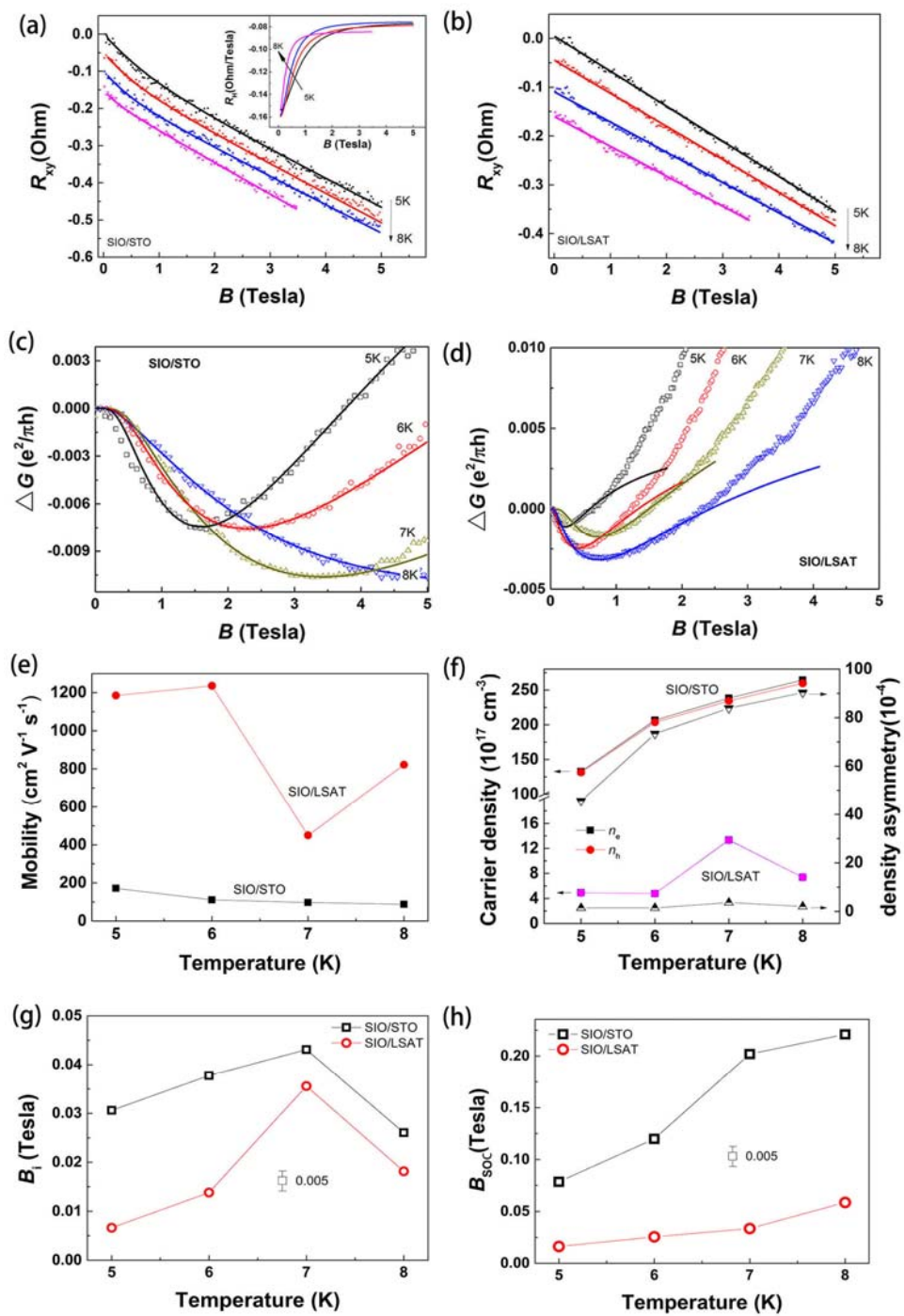

FIG. 3

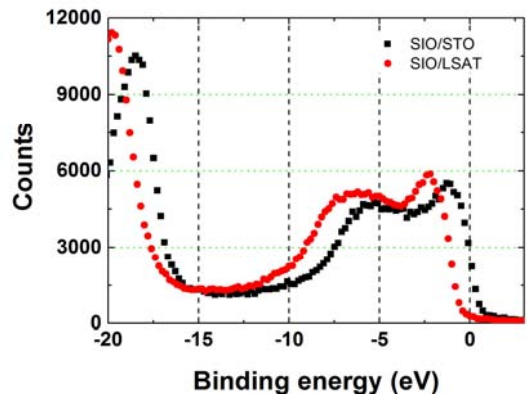

FIG. 4

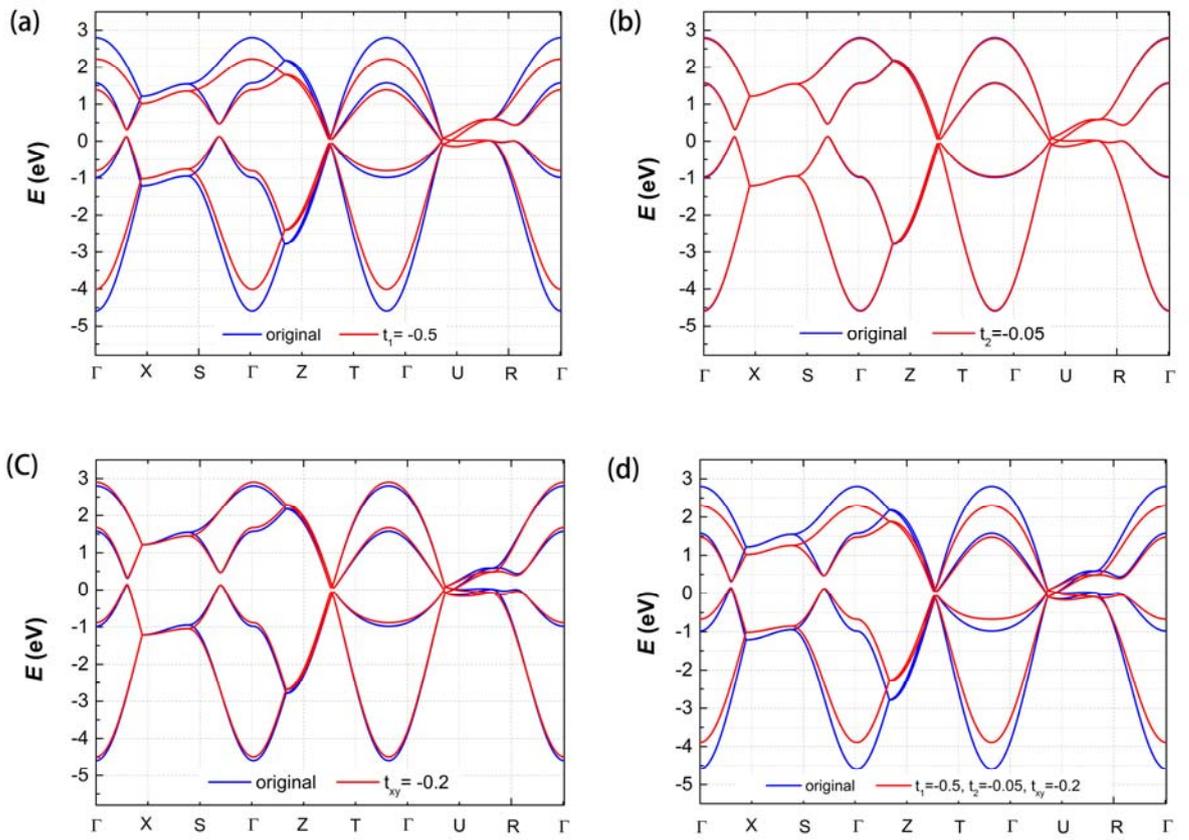

FIG. 5